\documentstyle[12pt]{article}

\newcommand{\ep}{\varepsilon}
\newcommand{\al}{\alpha}

\newcommand{\vp}{\varphi}
\newcommand{\be}{\begin{equation}}
\newcommand{\ee}{\end{equation}}
\newcommand{\ra}{\rightarrow}
\newcommand{\prt}{\partial}
\newcommand{\lt}{\left}
\newcommand{\rt}{\right}
\newcommand{\gze}{\stackrel{-}{g}(z,\ep )}
\newcommand{\Dt}{\Delta}
\newcommand{\sg}{\stackrel{-}{g}}
\newcommand{\sB}{\stackrel{\rightarrow}{B}}
\newcommand{\sS}{\stackrel{\rightarrow}{S}}
\newcommand{\sn}{\stackrel{\rightarrow}{n}}
\newcommand{\sr}{\stackrel{\rightarrow}{r}}
\newcommand{\sH}{\stackrel{\rightarrow}{H}}
\newcommand{\vt}{\vartheta}
\newcommand{\ga}{\gamma}
\newcommand{\om}{\omega}
\newcommand{\lgl}{\langle}
\newcommand{\rgl}{\rangle}
\newcommand{\Om}{\Omega}
\newcommand{\Ga}{\Gamma}

\begin{document}
\begin{center}
{\bf NONLINEAR SPIN DYNAMICS IN NUCLEAR MAGNETS \\ [5mm]
V.I.Yukalov} \\ [3mm]
{\it Department of Mathematics  and Statistics \\
Queen's University, Kingston, Ontario K7L 3N6, Canada \\
and \\
Bogolubov Laboratory of Theoretical Physics \\
Joint Institute for Nuclear Research, Dubna 141980, Russia}
\end{center}

\vspace{10cm}

{\bf PACS:} 76.20+q; 76.60Es

\newpage

\begin{abstract}

A method is developed for solving nonlinear systems of differential, or 
integrodifferential, equations with stochastic fields. The method makes it 
possible to give an accurate solution for an interesting physical problem: 
What are the peculiarities of nonlinear spin dynamics in nonequilibrium 
nuclear magnets coupled with a resonator? Evolution equations for nuclear 
spins are derived basing on a Hamiltonian with dipole interactions. The 
ensemble of spins is coupled with a resonator electric circuit. Seven 
types of main relaxation regimes are found: free induction, collective 
induction, free relaxation, collective relaxation, weak superradiance, pure 
superradiance, and triggered superradiance. The initial motion of spins can 
be originated by two reasons, either by an imposed initial coherence or by 
local spin fluctuations due to nonsecular dipole interactions. The relaxation 
regimes caused by the second reason cannot be described by the Bloch equations.
Numerical estimates show good agreement with experiment.

\end{abstract}

\newpage

{\Large{\bf I. Introduction}}

\vspace{1mm}

Spin dynamics in polarized nonequilibrium systems is usually described by 
using the Bloch equations for the components of uniform magnetization. The 
derivation of the Bloch equations from the evolution equations for a spin 
model can be found, for example, in ter Haar [1]. The solution of the Bloch 
equations for the case of small deviations from a stationary state is 
straightforward and well known in the theory of magnetic resonance [2]. The 
situation becomes more complicated when the spin system is coupled with a 
resonator. Then there appears an essential nonlinearity due to the action of 
resonator feedback field. The system of the coupled Bloch and resonator--field
equations is typical of the theory of maser amplifiers and generators [3].

The nonlinear system of the Bloch and resonator--field equations can be 
slightly simplified by invoking the slowly--varying amplitude approximation 
[3]. However, this does not help much, since the resulting equations are, 
as before, nonlinear. To achieve further simplification, one resorts to the 
adiabatic approximation which leads to the proportionality of the feedback 
field to transverse magnetization, that is, to the static coupling [4-7]. 
The adiabatic approximation, as is known [8], works well only at the final 
stage of relaxation processes when different variables adiabatically 
follow each other, but it cannot correctly describe intermediate stages 
where transient phenomena occur. The incorrectness of the static--coupling 
approximation is physically evident, as only moving spins, but not immovable, 
are able to induce a field in resonator. More accurate is the 
dynamic--coupling approximation [9] in which the feedback field is 
proportional to the time derivative of transverse magnetization. But both 
these, static--as well as dynamic--coupling, approximations do not take into 
account retardation effects that may be important for transient phenomena.

Moreover, the Bloch equations themselves may be inappropriate for explaining 
some kinds of relaxation processes. This concerns, for example, the 
interpretation of the recent series of experiments [10-15] observing nuclear 
spin superradiance. In these experiments a nonequilibrium system of polarized 
nuclear spins is placed inside a coil of a resonance electric circuit. The 
initial polarization is directed opposite to an external magnetic field. If 
this polarization is sufficiently high and the coupling with a resonator is 
enough strong, then the power of current, as a function of time, after some 
delay, displays a sharp burst with a damping time much shorter than the 
dephasing time $\;T_2\;$. This time behaviour of the current power is 
analogous to that of the radiation intensity of atoms or molecules in the 
case of optical superradiance. Because of this analogy, the corresponding 
coherent phenomenon in spin systems has also been called superradiance, or 
more concretely, spin superradiance. Friedberg and Hartmann [16] pointed out 
that the whole process of interaction of a spin system and a resonance coil, 
in fact, involves no radiation into free space but merely nonradiative 
transfer of energy from the sample to the coil, where the energy is dissipated
ohmically. Nevertheless, the term spin superradiance has become commonly used.
The excuse for this is not solely the formal analogy of temporal behaviour of 
current power, for spin systems, and of radiation intensity, for atomic and 
molecular systems, but also a deep physical similarity: The spin superradiance,
as well as optical superradiance, is a collective process of coherent 
self--organization. Although the self--organized coherence of spin motion 
develops not because of a common radiation field, as in atomic and molecular 
systems, but owing to a resonator feedback field. In addition, coherent motion
of spins inevitably produces coherent magnetodipole emission with properties 
completely analogous to superradiance of optical systems, though the 
magnetodipole radiation intensity is too weak to be measured as easy as 
the power of current [17].    

In the same way as for optical systems [18], one has to distinguish the pure 
from triggered spin superradiance. The {\it pure spin superradiance} is a 
purely self--organized process starting from an absolutely incoherent state 
when the average transverse magnetization is strictly zero. The {\it triggered
spin superradiance} is a process in which self--organization also plays an 
important role but whose beginning is triggered by an initial coherence 
imposed onto the spin system, that is by assuming that the mean transverse 
magnetization is not zero.

The interpretation of pure spin superradiance cannot be based on the Bloch 
equations because of the following. If the initial transverse magnetization 
is zero then, in the content of these equations, the relaxation of an inverted
spin system can be due only to two reasons: either to spin--lattice 
interactions characterized by a relaxation time $\;T_1\;$, or to thermal 
damping caused by the Nyquist noise of resonator. At very low temperature, 
typical of experiments [10-15] with polarized nuclear spins, the
spin--lattice relaxation time $\;T_1\;$ is much longer than the dephasing 
time $\;T_2\;$, therefore this mechanism cannot develop coherence. The 
resonator thermal damping, as is shown by Bloembergen and Pound [19], is 
negligibly small for macroscopic systems, the thermal relaxation time being 
proportional to the number of spins, $\;N\;$, in the sample, and so being 
much longer than not only $\;T_2\;$ but even $\;T_1\;$. Thus, the resonator 
Nyquist noise can never produce the initial thermal relaxation. The radiation 
field in the coil does not provide a microscopic thermal relaxation mechanism,
but the inhomogeneous internal, or local, fields are essential [19]. 
 
The Bloch equations cannot, in principle, describe the pure spin superradiance
and, in general, any other relaxation regimes in which no initial coherence is
imposed on the spin system. To treat all possible relaxation regimes for a 
nonequilibrium spin system, coupled with a resonator, it is necessary to take 
into account local spin fluctuations. This can be done by considering a 
microscopic model with realistic dipole interactions between nuclear spins. 
But, since the local fields are essential, we cannot invoke for a microscopic 
model a homogeneous approximation. The latter would immediately return us to 
the Bloch equations with the lost information on local spin fluctuations.

If the number of spins, $\;N\;$, is not too large, say between $\;10\;$ and 
$\;10^3\;$, then one can resort to a numerical solution of the corresponding 
evolution equations. Such a computer simulation, whose mathematical details 
can be found in [20], has been accomplished [17] and confirmed the crucial 
importance of local spin fluctuations. These are sufficient for describing 
the pure spin superradiance, with no influence of the resonator thermal 
noise.

Computer simulations, however, can give only a qualitative picture, as the 
number of spins involved is incomparably smaller than what one has in real 
samples with $\;N\;$ of the order of $\;10^{23}\;$. In addition, such 
simulations provide no analytical formulae making it very difficult, if 
possible, to classify all possible relaxation regimes occurring when varying 
the numerous parameters of the system.

The aim of the present paper is to untangle two mutually interrelated 
problems: first, to formulate a method  allowing an analytical solution 
for a system of nonlinear equations, with taking into account local 
fluctuating fields, as well as dynamic coupling and retardation effects; 
and second, to analyse various relaxation regimes of nonequilibrium nuclear 
magnets coupled with a resonator.

\vspace{3mm}

{\Large{\bf II. Method of Solution}}  

\vspace{1mm}

The method to be presented here may be used not only for the particular 
problem discussed in the Introduction, but for a wide variety of evolution 
equations for different systems. In this section we will preserve the 
generality of the presentation. All necessary specifications related to the 
spin dynamics in nuclear magnets will be expounded in the following sections. 
To better understand the principal ideas of the method, it is convenient to 
divide it into several steps.

\vspace{3mm}

{\bf 1. Separation of variables}

Suppose that in the problem under consideration there is a set
$$\ep =\{ \ep_i\; |i=1,2,\ldots ; \; |\ep_i| \ll 1\} $$
of small parameters. Depending on the way in which these parameters enter 
into the evolution equations, we may distinguish fast and slow variables. 
The terms describing local fluctuating fields can be treated as random, or 
stochastic, variables
$$ \vp = \{ \vp_i\; |i=1,2,\ldots ; \; \mu_\vp\} $$
with a probability measure $\;\mu_\vp\;$.

The fast variables
$$ u = \{ u_i(\vp ,t)\; |i=1,2,\ldots ; \; t\geq 0 \} $$
and slow variables
$$ s = \{ s_j(\vp ,t)\; |j=1,2,\ldots ; \; t\geq 0 \} $$
differ from each other by the properties of their evolution equations
\be
\frac{du}{dt} = f(u,s,\vp ,t,\ep )
\ee
and
\be
\frac{ds}{dt} =\ep g(u,s,\vp ,t,\ep ) ,
\ee
whose right--hand sides are such that the limit
\be
\lim_{\ep \ra 0}f(u,s,\vp ,t,\ep ) \neq 0
\ee
is not zero, while
\be
\lim_{\ep \ra 0}\ep g(u,s,\vp ,t,\ep ) = 0 .
\ee
Here and in what follows the matrix form of notation is used, according to 
which $\;f =\{ f_i\} ,\; g = \{ g_i\}\;$; and the product 
$\;\ep g =\{\sum_{j}c_{ij}\ep_jg_j\}\;$ is to be understood as a column of 
linear combinations with coefficients $\;c_{ij}\;$. All parameters, 
variables, functions, and coefficients can be complex except $\;t\geq 0\;$ 
representing time. The limit $\;\ep\ra 0\;$ means that all $\;\ep_i \ra 0\;$. 
The right--hand sides of (1) and (2) can contain integral operators, 
provided that the limits (3) and (4) hold. For brevity, the dependence of 
the fast, $\;u\;$, and slow, $\;s\;$, variables on the parameters $\;\ep\;$ 
is not explicitly written. Equations (1) and (2) are to be complimented by 
initial conditions
\be
u(\vp ,0) =u_0 , \qquad s(\vp ,0)=z_0 .
\ee
The limiting properties (3) and (4) explain why the evolution equations of 
the form (1) correspond to fast variables, as compared to the evolution 
equations of the type (2) describing slow variables.

The fast and slow variables are not necessarily simply defined for each given 
problem, but the aim of this step is to introduce such variables by using the 
information on the existence of small parameters and by choosing the 
appropriate changes of variables, so that finally they could be distinguished 
in the above sense. 

\vspace{2mm}

{\bf 2. Quasi--integrals of motion}

As far as the slow variables, by definition, vary with time much slower than 
the fast variables, the former may be considered as quasi--integrals of motion
for the latter. Then we can try to solve the equations for fast variables 
under slow variables kept as fixed parameters. With the notation 
\be
u = X , \qquad s = z ,
\ee
where $\;z\;$ is fixed, from (1) we have
\be
\frac{\prt X}{\prt t} = f(X,z,\vp ,t,\ep ) ,
\ee
which defines
\be
X = X(z,\vp ,t) .
\ee

The art of choosing variables is to get for (7) as simple equation as possible.
In many cases this can be done so that (7), under fixed $\;z\;$, becomes a 
system of linear equations. The quasi--integrals of motion play here a role 
similar to the guiding centers in the guiding--center approach [21].

\vspace{2mm}

{\bf 3. Method of averaging}

For the fast variable (8) we define the asymptotic period $\;T_0\;$ by the 
condition
\be
\lim_{\ep \ra 0}\lt | X (z,\vp ,t+T_0) - X(z,\vp ,t)\rt | = 0 .
\ee
If (9) gives several solutions for $\;T_0\;$, the smallest of them is to be 
taken. And if (9) has no solution for $\;T_0\;$, we put $\;T_0\ra\infty\;$.

To find the time evolution of quasi--integrals of motion, we substitute (8) 
into the right--hand side of (2) and introduce the averaged function
\be
\gze \equiv \int \lt [ \frac{1}{T_0} 
\int_{0}^{T_0} g\lt ( X(z,\vp ,t),z,\vp ,t,\ep \rt ) dt \rt ]d\mu_\vp .
\ee
Then the equation
\be
\frac{dz}{dt} =\ep\gze 
\ee
gives the sought time evolution.

The foundation for this step is the Krylov--Bogolubov method of averaging 
[22,23]. The major difference in our case is that the Krylov--Bogolubov 
vector field (10) is defined as an average with respect to time and, in 
addition, with respect to the stochastic variable $\;\vp\;$.

\vspace{2mm}

{\bf 4. Basic approximation}

The basic approximations for slow and fast variables are defined as follows. 
For the slow variables this is given by the solution
\be
z =z(t) 
\ee
of equation (11), with the initial condition
\be
z(0)=z_0 .
\ee
Substituting (12) into (8), we have
\be
x=x(\vp ,t) = X(z(t),\vp ,t)
\ee 
for fast variables. The integration constant appearing when solving (7) is to 
be found from the initial condition
\be
x(\vp ,0) =u_0 .
\ee

Note that (11) is,generally, a nonlinear equation, hence the basic 
approximations (12) and (14) take account of all nonlinearities essential for 
the considered dynamical process.

\vspace{2mm}

{\bf 5. Generalized expansion}

Corrections to the  basic approximation can be found by using the generalized 
asymptotic expansion,
$$ u =x(\vp ,t) +\sum_{n=1}^{\infty}x_n(\vp ,t)\ep^n , $$
\be
s =z(t) +\sum_{n=1}^{\infty}z_n(\vp ,t)\ep^n ,
\ee
about (12) and (14).

The right--hand sides of (1) and (2) are also to be expanded in a similar 
manner, as 
\be
f(u,s,\vp ,t,\ep ) = f(x,z,\vp ,t,\ep ) +
\sum_{n=1}^{\infty}f_n(\vp ,t,\ep )\ep^n  .
\ee
For example, in the first two orders we have
$$ f_1 =x_1f_x' +z_1f_z' , $$
$$ f_2 =x_2f_x' + z_2f_z' + x_1z_1f_{xz}'' +\frac{1}{2} \lt ( x_1^2f_{xx}'' +
z_1^2f_{zz}''\rt ) , $$
where the notation
$$ f_x' \equiv \frac{\prt}{\prt x}f(x,z,\vp ,t,\ep ); \qquad x=x(\vp ,t), \;
z=z(t) $$
is used.

The expansions (16) and (17) are to be substituted into the evolution 
equations (1) and (2). In doing this, we notice that, since because of (14)
$$ \frac{dx}{dt} = \lt ( \frac{\prt X}{\prt t}\rt )_z + 
\lt ( \frac{\prt X}{\prt z}\rt )_t\frac{dz}{dt} , $$
then invoking (7) and (11), we get
$$ \frac{dx}{dt} = f(x,z,\vp ,t,\ep ) + \ep \gze X_z'(\vp ,t) , $$
where
$$ X_z'(\vp ,t) \equiv \frac{\prt }{\prt z}X(z,\vp ,t); \qquad z=z(t) . $$

Equating similar terms with respect to the power of $\;\ep\;$, we obtain the 
equations for the corrections of arbitrary order. It is important to stress 
that all these equations are linear, thus, there is no principal difficulty 
in solving them. To exemplify this, at the same time avoiding cumbersome 
formulae, let us think of $\;\ep\;$ as of one parameter. Then for the 
first--order corrections we find the equations
$$ \frac{dx_1}{dt} = f_1(\vp ,t,\ep ) -\gze X_z'(\vp ,t) , $$
\be
\frac{dz_1}{dt} =g(x,z,\vp ,t,\ep ) -\gze .
\ee
The initial conditions, in compliance with (13) and (15), are 
\be
x_1(\vp ,0) =0, \qquad z_1(\vp ,0) =0.
\ee
For all subsequent orders we have
$$ \frac{dx_n}{dt} =f_n(\vp ,t,\ep ),  $$
\be
\frac{dz_n}{dt} = g_n(\vp ,t,\ep ), \qquad (n \geq 2),
\ee
with the initial conditions 
\be
x_n(\vp ,0)=0 , \qquad z_n(\vp ,0)=0 .
\ee

The first of Eqs.(18) can be reduced to the form
$$ \frac{dx_1}{dt} = x_1f_x' +\Dt_1 -\sg X_z' , $$ 
in which
$$ \Dt_1 \equiv f_1 - x_1f_x' = z_1f_z' . $$
As we see, the equation for $\;x_1\;$ is really linear, since 
\be
z_1(\vp ,t) =\int \lt [ g(x,z,\vp ,t,\ep ) -\gze \rt ] dt 
\ee
immediately follows from the second of Eqs.(18). The solution for this linear 
equation is
\be
x_1 =e^p\int e^{-p}\lt (\Dt_1 -\sg X_z'\rt ) dt ,
\ee
where
$$ p = p(\vp ,t,\ep ) \equiv \int f_x'(\vp ,t,\ep )dt . $$

For the second--order corrections, from (20), we find
\be
x_2 =e^p\int e^{-p}\Dt_2 dt , \qquad z_2 =\int g_1dt
\ee
with
$$ \Dt_2 \equiv f_2 - x_2f_x' = z_2f_z' + x_1z_1f_{xz}'' + 
\frac{1}{2} \lt ( x_1^2f_{xx}'' + z_1^2f_{zz}'' \rt ) . $$

Similarly, for the $\;n\;$--th order corrections we obtain the general 
formulae
$$ x_n = e^p\int e^{-p}\Dt_n dt ,  $$
\be
z_n = \int g_{n-1}dt , \qquad (n \geq 2),
\ee
in which
$$ \Dt_n \equiv f_n - x_nf_x' . $$

The  simplicity of obtaining the higher--order corrections, satisfying linear 
equations, is a considerable advantage of the suggested generalized asymptotic
expansion, as compared to the quiding--center approach [21] or averaging 
methods [22,23] in which each subsequent approximation order invokes more 
and more complicated nonlinear equations. Here we meet nonlinear equations 
only once, at the third step, when solving (11), which corresponds to the 
first--order averaging method.

The use of the averaging method only in one step makes it possible, from one 
side, to include all essential nonlinearity into our basic approximation and, 
from another side, to define all corrections by simple formulae. The idea of 
dividing solutions onto their principal parts, including essential 
nonlinearities, and perturbative corrections, defined by linear equations, 
greatly helps in solving complicated nonlinear problems [24]. This idea, 
actually, goes back to the Struble technique [25,26] imployed for solving the 
Mathieu equation. Note that the nonlinear principal part could be also defined
by other techniques known in the theory of singular perturbations [27], for 
instance, by using the methods of strained coordinates, multiple scales, 
nonlinear renormalizations, matched expansions, variation of parameters, and 
so on [28-30]. However, these  methods, as is discueed in [31,32], are more 
ambiguous, more cumbersome, and less general than the method of averaging.

Finally, we need to remember that, in our case, the solutions of nonlinear 
equations (1) and (2) contain the stochastic variable $\;\vp\;$. As far as 
observable quantities should not depend on that variable, this means that 
the former are to be averaged with respect to the random $\;\vp\;$ with a 
given probability measure. The solutions themselves are not necessary such 
quantities that can be measured directly, but usually, the observables are 
some functions or functionals of these solutions. This especially concerns 
the fast variables, while the slow variables are often directly measurable.

\vspace{3mm} 

{\Large{\bf III. Nuclear Magnet}}

\vspace{1mm}

The system of nuclear spins can be modeled, as is accepted in the theory of 
nuclear magnetic resonance [2], by the Hamiltonian
\be
\hat H =\frac{1}{2}\sum_{i\neq j}^{N}H_{ij} -\mu \sum_{i=1}^{N}\sB\sS_i
\ee
with the dipole interaction energy
\be
H_{ij} =\frac{\mu^2}{r^3_{ij}}\lt [ \sS_i\sS_j - 3\lt ( \sS_i\sn_{ij}\rt )
\lt ( \sS_j\sn_{ij}\rt )\rt ] ,
\ee
in which $\;\mu\;$ is a nuclear magneton, $\;\sS_i =\{ S_i^x,S_i^y,S_i^z\}\;$
is a spin operator, and
$$ \sn_{ij}\equiv \frac{\sr_{ij}}{r_{ij}}, \qquad 
\sr_{ij} \equiv \sr_i - \sr_j , \qquad r_{ij} \equiv \lt |\sr_{ij}\rt | . $$
The total magnetic field
\be
\sB = \sH_0 + \sH
\ee
consists of two parts,
\be
\sH_0 =H_0\stackrel{\ra}{e}_z , \qquad \sH =H\stackrel{\ra}{e}_x ; 
\ee
the first is an external magnetic field $\;H_0\;$ directed along the 
$\;z\;$--axis; the second, $\;H\;$, is a field of the coil of a resonance 
electric circuit, the coil axis being directed along the axis $\;x\;$. The 
sample is inserted into the coil.

Introduce the interactions
$$ a_{ij} \equiv \frac{\mu^2}{r_{ij}^3}\lt ( 1 - 3\cos^2\vt_{ij}\rt ) , $$
\be
b_{ij} \equiv -\frac{3\mu^2}{4r_{ij}^3}\sin^2\vt_{ij}\exp (-i2\vp_{ij}) ,
\ee
$$ c_{ij} \equiv -
\frac{3\mu^2}{4r_{ij}^3}\sin (2\vt_{ij})\exp (-i\vp_{ij}) , $$
in which $\;\vt_{ij}\;$ and $\;\vp_{ij}\;$ are the spherical angles of 
$\;\sn_{ij}\;$. These interactions have the symmetry property
\be
a_{ij} = a_{ji}, \qquad b_{ij}=b_{ji}, \qquad c_{ij}=c_{ji} .
\ee
Defining the ladder operators $\;S_i^-\;$ and $\;S_i^+\;$ by the expressions
$$ S_i^- = S_i^x -iS_i^y, \qquad S_i^+ = S_i + iS_i^y , $$
\be
S_j^x =\frac{1}{2}\lt ( S_i^- + S_i^+\rt ) , \qquad S_i^y =\frac{i}{2}\lt (
S_i^- - S_i^+\rt ) ,
\ee
and using (30), we may cast the dipole interaction energy (27) into the form
$$ H_{ij} = a_{ij}\lt ( S_i^zS_j^z -\frac{1}{2}S_i^+S_j^-\rt ) + 
b_{ij}S_i^+S_j^+ + b^*_{ij}S_i^-S_j^- + $$
\be
+ 2\lt ( c_{ij}S_i^+ +c_{ij}^*S_i^-\rt )S_j^z .
\ee

For the operators $\;S_i^-\;$ and $\;S_i^z\;$, satisfying the commutation 
relations
$$ \lt [ S_i^z,S_j^\pm\rt ] = \pm \delta_{ij}S_i^\pm , \qquad
\lt [ S_i^+,S_j^-\rt ] = 2\delta_{ij}S_i^z , $$
the Heisenberg equations of motion yield
$$ i\hbar \frac{d}{dt}S_i^- =
\sum_{j(\neq i)}^{N}\lt \{ a_{ij}\lt ( S_i^-S_j^z +
\frac{1}{2}S_i^zS_j^-\rt ) - 2b_{ij}S_i^zS_j^+ +\right. $$
\be
\left. +c_{ij}\lt ( S_i^-S_j^+ - 2S_i^zS_j^z\rt ) +
c_{ij}^*S_i^-S_j^-\rt \} - \mu H_0S_i^- +\mu HS_i^z
\ee
and
$$ i\hbar \frac{d}{dt}S_i^z =\sum_{j(\neq i)}^{N}\lt \{ \frac{a_{ij}}{4}
\lt ( S_i^-S_j^+ -S_i^+S_i^-\rt ) + b_{ij}S_i^+S_j^+ -b_{ij}^*S_i^-S_j^- +
\right. $$
\be
\left. +\lt ( c_{ij}S_i^+ -c_{ij}^*S_i^-\rt ) S_j^z\rt \} +
\frac{\mu}{2}H\lt ( S_i^- - S_i^+\rt ) -i\hbar\ga_1 
\lt ( S_i^z -\zeta_i\rt ) ,
\ee
where (35) is supplemented by a term taking into account spin--lattice 
interactions leading to the longitudinal damping $\;\ga_1\;$, and 
$\;\zeta_i\;$ being a stationary value of the spin $\;z\;$--component. 
The derivation of the spin--lattice term from microscopic spin--lattice 
interactions can be found in literature [1-3].

The initial state of the spin system is assumed to be nonequilibrium and 
characterized by a statistical operator $\;\hat\rho (0)\;$. So, the average 
spin
$$ \langle\sS_i\rangle \equiv {\rm Tr}\hat\rho (0)\sS_i(t) =
{\rm Tr} \hat\rho (t)\sS_i(0) $$
is a function of time. The evolution equations for averages can be obtained 
by using either the Liouville equation for the statistical operator 
$\;\hat\rho (t)\;$ or the Heisenberg equations of motion for operators. We 
prefer the latter way based on the Heisenberg equations (34) and (35).

\vspace{3mm}

{\Large{\bf IV. Resonator Field}}

\vspace{1mm}

The resonance electric circuit, coupled with the spin sample, is characterized 
by resistance $\;R\;$, inductance $\;L\;$ and capacity $\;C\;$. The coil, in 
which the sample is immersed, has $\;n\;$ turns of cross section $\;A_0\;$ 
over a length $\;l\;$. The magnetic field inside the coil,
\be 
H =\frac{4\pi n}{cl}j ,
\ee
is formed by an electric current satisfying the Kirchhoff equation
\be
L\frac{dj}{dt} +Rj +\frac{1}{C}\int_{0}^{t}j(\tau )d\tau = 
-\frac{d\Phi}{dt} +E_f ,
\ee
in which $\;E_f\;$ is an electromotive force of external fields, if any, and 
of the thermal Nyquist noise; the magnetic flux
\be
\Phi = \frac{4\pi}{c}nA_0\eta\rho M_x 
\ee
is due to the $\;x\;$--component of the magnetization
\be
M_x =\frac{\mu}{N}\sum_{i=1}^N\langle S_i^x\rangle ;
\ee
and the filling factor $\;\eta\;$ and spin density $\;\rho\;$ are
$$ \eta \equiv \frac{V}{V_0}, \qquad \rho \equiv \frac{N}{V} \qquad
(V_0 \equiv lA_0) , $$
respectively.

The resonance electric circuit will be called, for brevity, the resonator, and
the internal coil field (36), the resonator field. For the latter, the 
Kirchhoff equation (37) can be rewritten as
\be
\frac{dH}{dt} + 2\ga_3H +\om^2\int_{0}^{t}H(\tau )d\tau = 
-4\pi\eta\rho\frac{dM_x}{dt} + \frac{cE_f}{nA_0} ,
\ee
where
$$ \om \equiv \frac{1}{\sqrt{LC}} \qquad 
\lt ( L\equiv \frac{4\pi n^2A_0}{c^2l}\rt ) $$
is the resonator natural frequency, and
$$ \ga_3 \equiv \frac{R}{2L} =\frac{\om}{2Q} \qquad 
\lt ( Q\equiv \frac{\om L}{R}\rt ) $$
is the resonator damping.

It is convenient to introduce the dimensionless resonator field
\be
h \equiv\frac{\mu H}{\hbar\ga_3} ,
\ee
driving force
\be
f \equiv \frac{c\mu E_f}{nA_0\hbar\ga_3^2},
\ee
and the dimensionless average magnetization
\be
s_\nu \equiv\frac{M_\nu}{\mu} = 
\frac{1}{N}\sum_{i=1}^{N}\langle S_i^\nu\rangle ,
\ee
in which $\;\nu =x,y,z\;$. Define the coupling constant
\be
\al_0 \equiv\pi\eta\frac{\rho\mu^2}{\hbar\ga_3} ,
\ee
characterizing the strength of coupling between the spin system and 
resonator. Then the Kirchhoff equation (40) acquires the form
\be
\frac{dh}{dt} +2\ga_3h +\om^2\int_{0}^{t}h(\tau )d\tau = 
-4\al_0\frac{ds_x}{dt} +\ga_3f .
\ee
The resonator field $\;h\;$, as is seen from (45), can be induced by a 
driving force $\;f\;$ and by moving, but not static, transverse 
magnetization.

\vspace{3mm}

{\Large{\bf V. Average Magnetization}}

\vspace{1mm}

The statistical averaging of a spin operator $\;S_i^\al = S_i^\al (t)\;$, 
with $\;\al = x,y,z\;$, is given by
\be
\lgl S_i^\al\rgl \equiv {\rm Tr}\hat\rho (0) S_i^\al (t) .
\ee
We shall use the notation
\be
u_i \equiv \lgl S_i^-\rgl , \qquad s_i \equiv \lgl S_i^z\rgl .
\ee
The statistical operator $\;\hat\rho (0)\;$ in (46) defines the initial values
of (47), that is, $\;u_i(0)\;$ and $\;s_i(0)\;$.

To obtain the evolution equations for the transverse, $\;u_i\;$, and 
longitudinal, $\;s_i\;$, magnetizations, we have to average the equations of 
motion (34) and (35), according to (46). The dipole interactions are of 
long--range type, therefore the double spin correlations can be decoupled 
in the mean--field approximation
$$ \lgl S_i^\al S_j^\beta\rgl \ra \lgl S_i^\al\rgl\lgl S_j^\beta\rgl \qquad
(i\neq j) . $$
Although this decoupling is well justified for long--range forces [33], it 
has a deficiency that is important for nonequilibrium processes: it does not 
take into account the attenuation due to spin--spin interactions. This 
attenuation appears in the higher--order corrections to the mean--field 
approximation. The derivation of the spin--spin damping $\;\ga_2\;$ in the 
second--order perturbation theory can be found e.g. in ter Haar [1]. This 
damping has to be retained for a correct description of relaxation process, 
though $\;\ga_2\;$ is much smaller than the Larmor frequency
\be
\om_0 \equiv\frac{\mu H_0}{\hbar} > 0 .
\ee
At the same time the small second--order corrections to the oscillation 
frequency (48) can be neglected; alternatively, they can be included into 
the definition of $\;\om_0\;$. The mean--field decoupling with corrections 
leading to the appearance of the spin--spin relaxation parameter $\;\ga_2\;$ 
can be called the corrected mean--field approximation. Within the framework 
of this approximation, the averaging of (34) and (35) yields for the 
variables in (47) the equations
$$ i\frac{du_i}{dt} = - (\om_0 +i\ga_2)u_i +\ga_3hs_i + $$
\be
+\frac{1}{\hbar}\sum_{j(\neq i)}^{N}\lt \{ \frac{a_{ij}}{2}\lt ( s_iu_j +
2u_is_j\rt ) - 2b_{ij}s_iu_j^* + c_{ij} ( u_iu_j^* - 2s_is_j ) + 
c_{ij}^*u_iu_j\rt \}
\ee
and 
$$ i\frac{ds_i}{dt} = \frac{1}{2}\ga_3h\lt ( u_i - u_i^*\rt ) + 
 \frac{1}{\hbar}\sum_{j(\neq i)}^{N}\lt \{ a_{ij}(u_iu_j^* - u_i^*u_j) +
b_{ij}u_i^*u_j^* - b_{ij}^*u_iu_j + \right. $$
\be
\left. + (c_{ij}u_i^* - c_{ij}^*u_i)s_j\rt \} - i\ga_1(s -\zeta_i) .
\ee

Introduce the  arithmetic averages
\be
u \equiv \frac{1}{N}\sum_{i=1}^{N}u_i , \qquad s \equiv 
\frac{1}{N}\sum_{i=1}^{N}s_i 
\ee
for the transverse and longitudinal magnetizations, respectively, and also for
a stationary magnetization
\be
\zeta \equiv \frac{1}{N}\sum_{i=1}^{N}\zeta_i .
\ee
Define
\be
\delta_i \equiv \frac{1}{N}\sum_{j(\neq i)}^{N}\lt ( \frac{3}{2}a_{ij}s_j + 
c_{ij}u_j^* + c_{ij}^*u_j\rt )
\ee
which is a real quantity, and
\be
\vp_i \equiv -\frac{2}{\hbar}\sum_{j(\neq i)} \lt ( b_{ij}u_j^* + 
c_{ij}s_j\rt )
\ee which is complex.

For the averages in (51), from (49) and (50), using the symmetry property 
(31), we find 
\be
i\frac{du}{dt} = - (\om_0 +i\ga_2 ) u + \ga_3hs + \frac{1}{N}
\sum_{i=1}^{N}(\delta_iu_i +\vp_is_i)
\ee 
and
\be
i\frac{ds}{dt} = \frac{1}{2}\ga_3h(u - u^*) - i\ga_1(s -\zeta ) +
\frac{1}{2N}\sum_{i=1}^{N}(\vp_i^*u_i - \vp_iu_i^*) .
\ee

The quantities (53) and (54) are local fluctuating fields [1], whose existence
is due to the inhomogeneity of spin distribution. If one would resort to a 
homogeneous approximation, in which $\;u_j\;$ and $\;s_j\;$ do not depend on 
the index $\;j\;$, then $\;\delta_i\;$ and $\;\vp_i\;$ would be zero, since 
for the dipole interactions (30) we have
$$ \sum_{j(\neq i)}^{N}a_{ij} \simeq \sum_{j(\neq i)}^{N}b_{ij} \simeq
\sum_{j(\neq i)}^{N}c_{ij} \simeq 0 $$
when $\;N\ra\infty\;$ and the spin sample is macroscopic in all three 
dimensions. The above sums can be nonzero if the number of spins is not high
$\;(N<10)\;$ or if the sample has a specially prepared irregular shape. Then 
the nonzero values of these sums are defined by a nonuniformity in the space 
distribution of spins in the vicinity of the sample surface. Such a boundary 
nonuniformity for small, at least in one of dimensions, samples can lead to 
unisotropic effects in relaxation processes [16,34]. This kind of 
inhomogeneity of a sample inside a coil can be explicitly taken into account 
in the definition of the effective factor [19].

It is worth emphasizing that even when the spin sample is macroscopic and has 
a regular shape, so that the above sums over the dipole interactions (30) are 
nullified, nevertheless, the local fields (53) and (54) are nonzero if one 
does not invoke a uniform approximation for the magnetizations $\;u_j\;$ and 
$\;s_j\;$. The local nonuniformities contribute to the inhomogeneous dipole 
broadening [35]. What is the most important is that without taking into 
account such local fluctuating fields it is impossible, as has been stressed 
by Bloembergen and Pound [19], to provide a  correct description of relaxation
in spin systems.

At the same time, if (53) and (54) depend on the index $\;i\;$ showing their 
local position, then the equations (55) and (56) are not closed, but for the 
case of $\;N\;$ spins we need to deal with a system of $\;3N\;$ equations 
defined in (49) and (50). For a macroscopic sample with $\;N\sim10^{23}\;$, 
to deal with such a number of nonlinear differential equations is a task that 
is not affordable even for a computer.

A way out of this trouble is as follows. We may treat (53) and (54) as random 
fluctuating fields with a distribution given by a probability measure 
$\;\mu_\vp\;$. That is, we may put into correspondence to the local fields 
(53) and (54) stochastic fields
$$ \{\vp_0\} \leftrightarrow \{\delta_i\} , \qquad \{\vp\} \leftrightarrow
\{\vp_i\} , $$
in which $\;\vp_0\;$ is real, representing the real $\;\delta_i\;$, and 
$\;\vp\;$ is complex representing the complex $\;\vp_i\;$. At the present 
stage an explicit form of the probability measure $\;\mu_\vp\;$ is not 
important and will be considered later.

With the stochastic representation of local fields in mind, equations (55) 
and (56) are reduced to
\be
\frac{du}{dt} = i(\om_0 -\vp_0 +i\ga_2)u - i(\ga_3h +\vp )s
\ee
and
\be
\frac{ds}{dt} =\frac{i}{2}\lt ( \ga_3h+\vp \rt ) u^* -
\frac{i}{2}\lt (\ga_3h +
\vp^*\rt ) u -\ga_1 (s-\zeta ) .
\ee
Since $\;u\;$ is complex, the third equation, additional to (57) and (58), 
can be the equation for $\;u^*\;$ or for $\;|u|^2\;$. For the latter we have
\be
\frac{d}{dt}|u|^2 = -2\ga_2|u|^2 - i(\ga_3h +\vp )su^* + i(\ga_3h +\vp^*)su .
\ee
These equations are to be complimented by initial conditions
\be
u(0)=u_0 , \qquad s(0)=z_0 .
\ee

Eqs.(57)-(59) for the magnetizations plus Eq.(45) for the resonator field form
the basic system of equations permitting a correct description of relaxation 
processes for a spin sample coupled with a resonator. The physical meaning of 
all terms in these equations is quite transparent: The real random field 
$\;\vp_0\;$ shifts the oscillation frequency; and the term $\;\ga_3h+\vp\;$ 
plays the role of an effective field acting on spins, $\;h\;$ being the 
resonator field and $\;\vp\;$, stochastic field caused by local fluctuations.
If in (57)-(59) we would put $\;\vp_0\;$ and $\;\vp\;$ zero, then we would 
return to the Bloch equations; however the presence of these random fields, 
as is discussed above and will be demonstrated in what follows, provides a 
crucial relaxation mechanism. Note that the stochastic local fields 
interconnect the transverse and longitudinal components of magnetization, 
but do not change the absolute value of the latter whose time variation
$$ \frac{d}{dt}\lt ( |u|^2 +s^2\rt ) = - 2\ga_2|u|^2 -2\ga_1s(s-\zeta ) $$
is caused only by the spin--spin dephasing collisions and spin--lattice 
interactions.

If we would decide to invoke the adiabatic approximation, in the way one 
usually does, then we should put $\;\frac{du}{dt} \ra 0\;$ in (57) which 
immediately results in the linear relation between $\;h\;$ and $\;u\;$, 
that is, in the static approximation. However, as is discussed in the 
Introduction, such an approximation could be reasonable only at the final 
stage of relaxation, but cannot correctly describe transient phenomena.

\vspace{3mm}

{\Large{\bf VI. Separation of Variables}}

\vspace{1mm}

To solve the system of equations (57)-(59) and (45), we use the method 
developed in Sec.II. To this end, we need to separate fast from slow variables
by defining the appropriate small parameters. Usually, the widths $\;\ga_1\;$ 
and $\;\ga_2\;$ are small as compared to $\;\om_0\;$; and $\;\ga_3\;$ is small
as compared to $\;\om\;$. The stochastic fields $\;\vp_0\;$ and $\;\vp\;$ are 
also to be considered as small, since the corresponding local fields (53) and 
(54), as is evident from their definition, are of the order of the local 
dipole interactions, that is, of the order of $\;\ga_*\;$ which is a part of 
the inhomogeneous dipole broadening; $\;\ga_*\;$ being much smaller than 
$\;\om_0\;$. Thus, there are four small parameters:
\be
\frac{\ga_1}{\om_0}\ll 1, \qquad \frac{\ga_2}{\om_0} \ll 1, \qquad 
\frac{\ga_*}{\om_0} \ll 1, \qquad \frac{\ga_3}{\om} \ll 1 .
\ee
An additional small parameter appears in the quasiresonance situation when 
the resonator natural frequency is close to the Larmor frequency of spins. 
Then the detuning from the resonance, $\;\Dt\;$, gives another small 
parameter
\be
\frac{|\Dt|}{\om_0} \ll 1 \qquad (\Dt \equiv \om -\om_0 ) .
\ee
 
The quantities inverse to the corresponding widths define the characteristic 
times
\be
T_1 \equiv \frac{1}{\ga_1}, \qquad T_2 \equiv \frac{1}{\ga_2} , \qquad 
T_2^* \equiv \frac{1}{\ga_*}, \qquad T_3 \equiv \frac{1}{\ga_3} ,
\ee
among which $\;T_1\;$ is the spin--lattice relaxation time; $\;T_2\;$, 
spin--spin dephasing time; $\;T_2^*\;$, inhomogeneous dephasing time; 
$\;T_3\;$, resonator ringing time. To be more cautions, it is worth noting 
that, in our case, the width $\;\ga_*\;$ is due to local spin fluctuations 
which is only one of the possible mechanisms of inhomogeneous broadening. The 
latter arises also owing to crystalline defects, hyperfine interactions and 
other inhomogeneities [35] that are not included in our consideration. 
Therefore, here $\;T_2^*\;$ is of the order of $\;T_2\;$, both of them being 
related to dipole interactions, so $\;\ga_*\sim\ga_2\;$. The existence of the 
small parameters (61) means that the oscillation period
\be
T_0 \equiv \frac{2\pi}{\om_0} \ll {\rm min}\{ T_1,T_2,T_2^*,T_3\}
\ee
is the shortest time as compared to the characteristic times (63).

To check the properties (3) and (4), we have to take the limit in Eqs.(45) and
(57)-(59) by putting zero all small parameters (61) and (62), and respectively,
$\;\vp_0\;$ and $\;\vp\;$. This procedure yields the limits
$$ \frac{du}{dt} \ra i\om_0u , $$
\be
\frac{dh}{dt}\ra -\om^2\int_{0}^{t}h(\tau )d\tau - 2i\al_0\om_0(u-u^*) ,
\ee
$$ \frac{ds}{dt}\ra 0 , \qquad \frac{d}{dt}|u|^2 \ra 0 , $$
which shows that $\;u\;$ and $\;h\;$ are to be treated as fast, while $\;s\;$ 
and $\;|u|^2\;$ as slow variables. The first of limits in (65) also shows that
the adiabatic approximation is not appropriate when $\;u\;$ is not zero.

At the next step we have to consider the slow variables as quasi--integrals 
of motion for fast variables. The corresponding equations (57) and (45), with 
the notation
\be
u=x-iy , \qquad s=z ,
\ee
where $\;z\;$ is kept as a fixed parameter, can be written in the form
$$ \frac{dx}{dt} =-\ga_2x + \om_\vp y -\vp_2z , $$
\be
\frac{dy}{dt} = -\om_\vp x -\ga_2y + (\ga_3h + \vp_1)z , 
\ee
$$ \frac{dh}{dt}= -2\ga_3h - \om^2\int_{0}^{t}h(\tau )d\tau -
4\al_0\frac{dx}{dt}+\ga_3f , $$
in which
\be
\om_\vp \equiv \om_0 -\vp_0
\ee
is the shifted frequency, and the stochastic field
\be
\vp =\vp_1-i\vp_2
\ee
is separated into its real and imaginary parts. The initial conditions to (67)
are
\be
x(0)=x_0, \qquad y(0)=y_0, \qquad h(0)=0 .
\ee

It is remarkable that the system of three integro--differential equations (67),
under fixed $\;z\;$, is linear, thus can be solved exactly by imploying, 
e.g., the method of the Laplace transforms. Equivalently, differentiating 
the last of the equations in (67), we may convert (67) into a linear system 
of five ordinary differential equations, which is again exactly solvable by 
means of either the method of the Laplace transforms or the matrix methods.

The exact solution of (67) is so cumbersome that it is not pleasure to write 
it down explicitly. Fortunately, we can simplify it by using the existence of 
the small parameters (61) and (62). Such a simplification can be done directly
by, first, finding an exact solution of (67) and, second, performing some 
expansions in small parameters. However, this direct way is extremely tedious 
and does not provide an insight into the physics of the made simplifications. 
The same final result can be obtained in another way which is much less 
wearisome and more physically clear, and which is explained below.

The formal solution of the last equation in (67) can be written as the sum
\be
h=h_s+h_f ,
\ee
in which the first term is a feedback field induced in the resonator by moving 
spins and the second term is a resonator field formed by driving forces. The 
resonator feedback field may be presented either as the convolution
\be
h_s =-4\al_0\int_{0}^{t}\frac{d}{dt}x(t-\tau)W(\tau)d\tau
\ee
or as the Stieltjes integral
$$ h_s=-4\al_0\int_{0}^{t}W(t-\tau)dx(\tau) , $$
and the resonator forcing field is given by the convolution
\be
h_f =\ga_3\int_{0}^{t}W(t-\tau)f(\tau)d\tau ,
\ee
where the transfer function is
\be
W(t)=\lt ( \cos\om_3t-\frac{\ga_3}{\om_3}\sin\om_3t\rt )e^{-\ga_3t} 
\ee
with 
$$ \om_3\equiv \sqrt{\om^2-\ga_3^2} . $$

The action of the resonator field (71) on the spin system involves, as follows
from (67), the small parameter $\;\ga_3\;$. Neglecting this parameter reduces 
the first two equations in (67) to
$$ \frac{dx}{dt}\cong -\ga_2x +\om_\vp y -\vp_2z , $$
\be
\frac{dy}{dt} \cong -\om_\vp x -\ga_2y +\vp_1z .
\ee
The solution to (75) is
$$ x\cong \lt ( a_0\cos\om_\vp t +b_0\sin\om_\vp t\rt )e^{-\ga_2t} +
\frac{\vp_1}{\om_\vp}z , $$
\be
y \cong \lt ( b_0\cos\om_\vp t - a_0\sin\om_\vp t\rt )e^{-\ga_2t} +
\frac{\vp_2}{\om_\vp}z ,
\ee
where
$$ a_0 = x_0 -\frac{\vp_1}{\om_\vp}z , \qquad 
b_0 = y_0 -\frac{\vp_2}{\om_\vp}z . $$
Imploying (76) in (72) gives the feedback field
\be
h_s =-\frac{2}{\ga_3}\lt [ \al_1\frac{dx}{dt} +
\al_2\om_\vp \lt ( x -\frac{\vp_1}{\om_\vp}z\rt )\rt ] ,
\ee
in which
$$ \al_1 =\frac{\al_0\ga_3(\ga_2 -\ga_3)}{(\ga_2-\ga_3)^2 +(\Dt +\vp_0)^2}
\lt [ e^{(\ga_2-\ga_3)t} -1 \rt ] , $$
\be
\al_2 =\frac{\al_0\ga_3(\Dt +\vp_0)}{(\ga_2-\ga_3)^2+(\Dt+\vp_0)^2}
\lt [ e^{(\ga_2 -\ga_3)t} -1\rt ].
\ee

If in the expression (77) we put $\;\al_1 =0,\;\al_2=const\;$, we return to 
the static--coupling approximation, while if we put 
$\;\al_1=const,\;\al_2=0\;$, then we get the dynamic--coupling approximation 
[9]. However, in general, $\;\al_1=\al_1(t)\;$ and $\;\al_2=\al_2(t)\;$ are 
nonzero functions of time. The temporal dependence of the coupling functions 
in (78) portrays the retardation due to a gradual switching on of the coupling
between the spins and resonator. Really, as is seen from (78), at the initial 
moment the coupling is absent
$$ \al_1(0)=\al_2(0)=0 . $$

Using the first of the equations in (67) for (77) yields
\be
h_s =\frac{2}{\ga_3}\lt [ (\al_1\ga_2 -\al_2\om_\vp )x -\al_1\om_\vp y +
(\al_1\vp_2 +\al_2\vp_1)z\rt ] .
\ee
Substituting (79) back into (67) reduces the system of three 
integro--differential equations to the system of two ordinary differential 
equations
$$ \frac{dx}{dt} = -\ga_2x +\om_\vp y -\vp_2z , $$
$$ \frac{dy}{dt} = -(\om_\vp - 2\al_1\ga_2z + 2\al_2\om_\vp z)x -
(\ga_2 +2\al_1\om_\vp z)y + $$
\be
+(\vp_1 +2\al_1\vp_2z +2\al_2\vp_1z+\ga_3h_f)z
\ee
for the fast variables.

\vspace{3mm}

{\Large{\bf VII. Fast Variables}}

\vspace{1mm}

There is no problem in solving (80), which gives 
$$ x =\lt ( a_1\cos\Om_\vp t + b_1\sin\Om_\vp t\rt )e^{-\Ga_\vp t} +
x_\vp +x_f , $$
\be
y = \lt ( a_2\cos\Om_\vp t + b_2\sin\Om_\vp t\rt ) e^{-\Ga_\vp t} + y_\vp 
+ y_f ,
\ee
where the first parts describe the spin oscillations with the effective 
frequency
\be
\Om_\vp =\om_\vp\lt ( 1 -\al_1^2z^2 +2\al_2z\rt )^{1/2} ,
\ee
effective attenuation
\be
\Ga_\vp =\ga_2 +\al_1z\om_\vp ,
\ee
and coefficients
$$ a_1 =x_0 -x_\vp , \qquad a_2 = y_0 - y_\vp , $$
$$ b_1 =\frac{\om_\vp}{\Om_\vp} \lt ( y_0 +\al_1zx_0\rt ) -
\frac{\om_\vp z}{\Om_\vp^2+\Ga_\vp^2}\lt [ \lt ( 1 +2\al_2z\rt ) 
\frac{\Ga_\vp}{\Om_\vp}\vp_1 + \lt ( \frac{\Om_\vp}{\om_\vp} + 
\al_1z\frac{\Ga_\vp}{\Om_\vp}\rt )\vp_2\rt ] , $$
$$ b_2 =-\frac{\om_\vp}{\Om_\vp}\lt [\lt ( 1+2\al_2z\rt )x_0 +\al_1zy_0\rt ]+ 
 \frac{\om_\vp z}{\Om_\vp^2+\Ga_\vp^2}\lt \{\lt ( 1 +2\al_2z\rt )
\lt (\frac{\Om_\vp}{\om_\vp}+\al_1z\frac{\Ga_\vp}{\Om_\vp}\rt )\vp_1 +
\right. $$
$$ \left. +\lt [ 2\al_1z\lt ( \frac{\Om_\vp}{\om_\vp} +\al_1z
\frac{\Ga_\vp}{\Om_\vp}\rt ) - \lt ( 1 +2\al_2z\rt ) 
\frac{\Ga_\vp}{\Om_\vp}\rt ]\vp_2\rt \} ; $$
the terms
$$ x_\vp =\frac{\om_\vp z}{\Om_\vp^2 +\Ga_\vp^2}\lt [ ( 1 +2\al_2z)\vp_1 -
\lt ( \frac{\Ga_\vp}{\om_\vp} -\al_1z\rt )\vp_2\rt ] , $$
$$ y_\vp =\frac{\om_\vp z}{\Om_\vp^2 +\Ga_\vp^2}\lt \{ \lt ( 1 +2\al_2z\rt )
\lt (\frac{\Ga_\vp}{\om_\vp} -\al_1z\rt )\vp_1 + \right. $$
\be
\left. + \lt [ 1 +2\al_2z +2\al_1z\lt (\frac{\Ga_\vp}{\om_\vp} -
\al_1z\rt )\rt ]\vp_2\rt \}
\ee
are originated by the local random fields; and the last terms
$$ x_f =\ga_3\int_{0}^{t}G_1(t-\tau)h_f(\tau)d\tau , $$
\be
y_f =\ga_3\int_{0}^{t}G_2(t-\tau)h_f(\tau)d\tau
\ee
are due to the resonator forcing field; the Green functions being
$$ G_1(t) = z\frac{\om_\vp}{\Om_\vp}\sin\Om_\vp t\cdot e^{-\Ga_\vp t} , $$
$$ G_2(t) = z\cos\Om_\vp t\cdot e^{-\Ga_\vp t} -\al_1zG_1(t) . $$

In this way, the fast variable $\;u\;$, defined by Eq.(57), becomes
\be
u=u_s + u_\vp + u_f ,
\ee
where 
$$ u_s =\lt ( c_1e^{i\Om_\vp t} + c_2e^{-i\Om_\vp t}\rt ) e^{-\Ga_\vp t} , $$
\be
u_\vp = x_\vp -iy_\vp , \qquad u_f=x_f -iy_f ,
\ee
and
$$ c_1 =\frac{1}{2}\lt ( a_1 - b_2 \rt ) -\frac{i}{2}\lt ( b_1 + a_2\rt ) , $$
$$ c_2 =\frac{1}{2}\lt ( a_1 + b_2 \rt ) +\frac{i}{2}\lt ( b_1 - a_2\rt ) . $$

To find an explicit expression for $\;u_f\;$, induced by an electromotive 
force $\;E_f\;$, entering into the right--hand side of the Kirchhoff equation
(37), we need to concretize the form of $\;E_f\;$. Accepting for the latter 
the standard expression
\be
E_f =E_0\cos\om t ,
\ee
for the driving force (42) we have
\be
f = f_0\cos\om t; \qquad f_0 \equiv\frac{c\mu E_0}{nA_0\hbar\ga_3^2}.
\ee
Then the convolution (73), with the transfer function (74), gives
\be
h_f =\frac{f_0}{2}\lt ( \cos\om t -\frac{\ga_3}{\om}\sin\om t\rt )
\lt ( 1 - e^{-\ga_3t}\rt ) .
\ee
Substituting the resonator forcing field (90) into (85), we get
$$ x_f=\lt ( f_1e^{i\om t}+f_1^*e^{-i\om t}\rt )\lt ( 1 - e^{-\ga_3t}\rt ), $$
\be
y_f =\lt ( f_2e^{i\om t} + f_2^*e^{-i\om t}\rt )\lt ( 1 -e^{-\ga_3t}\rt ),
\ee
where the coefficients are
$$ f_1 = -\frac{f_0\om_\vp\ga_3z}{8\Om_\vp (\Dt_\vp^2 +\Ga_\vp^2)}
\lt ( \Dt_\vp + i\Ga_\vp\rt ) , $$
$$ f_2 = f_1\lt ( i\frac{\Om_\vp}{\om_\vp} -\al_1z\rt ) , $$
and the effective detuning is
\be
\Dt_\vp \equiv \om -\Om_\vp .
\ee
Therefore $\;u_f\;$ in (87) becomes
\be
u_f =\lt ( d_1e^{i\om t} + d_2e^{-i\om t}\rt ) \lt ( 1 - e^{-\ga_3t}\rt )
\ee
with the coefficients
$$ d_1 = f_1\lt ( 1 + \frac{\Om_\vp}{\om_\vp} + i\al_1z\rt ) , $$
$$ d_2 = f_1^*\lt ( 1 -\frac{\Om_\vp}{\om_\vp} + i\al_1z\rt ) . $$

Finally, the fast variable $\;h\;$, given by the sum (71), is composed of the 
terms (79) and (90) for which we have
\be
h_s=\frac{\om_\vp}{\ga_3}\lt [ i(\al_1 +i\al_2)u^* - i(\al_1 -i\al_2)u +
\frac{2}{\om_\vp}\lt ( \al_1\vp_2 +\al_2\vp_1\rt ) z\rt ]
\ee
and
\be
h_f=\frac{f_0}{4}\lt ( 1-i\frac{\ga_3}{\om}\rt )
\lt ( e^{i\om t} + e^{-i\om t}\rt )\lt ( 1 -e^{-\ga_3t}\rt ) .
\ee
The factors $\;(1-e^{-\ga_3t})\;$ in (90), (91),(93), and (95) describe the 
retardation in the interaction of the sample and resonator.

\vspace{3mm}

{\Large{\bf VIII. Slow Variables}}

\vspace{1mm}

At the next step of the method, displayed in Sec.II, we have to substitute 
the fast variables (86), (94), and (95) into the equations (58) and (59) for 
the slow variables
\be
s=z, \qquad |u| =v,
\ee
averaging the right--hand sides of (58) and (59) over the asymptotic period of
fast oscillations and also over a distribution of stochastic fields 
characterized by a probability measure $\;\mu_\vp\;$. The asymptotic period, 
according to the definition (9), is just (64). Let us denote the double 
averaging of a function $\;F=F_\vp(t)\;$, over the asymptotic period and over 
stochastic fields, as
\be
\lgl\lgl F\rgl\rgl \equiv \int\lt [\frac{1}{T_0}\int_{0}^{T_0} F_\vp(t)dt\rt ]
d\mu_\vp .
\ee
Since $\;\vp_0\;$ is real and $\;\vp=\vp_1-i\vp_2\;$ is complex, there are 
three independent real components of the stochastic fields, thence the 
differential measure $\;d\mu_\vp\;$ can be written as the product
$$ d\mu_\vp =d\mu(\vp_0)d\mu(\vp_1)d\mu(\vp_2) . $$
It is customary to model the distribution of local dipole fields in spin 
systems by a Gaussian distribution [3,35]. Accepting this and assuming, for 
simplicity, that each distribution of $\;\vp_\nu\;$, with $\;\nu=0,1,2\;$, 
has the same width $\;\ga_*\;$, we get
$$ d\mu(\vp_\nu) =\frac{1}{\sqrt{2\pi}}\exp\lt \{-\frac{1}{2}\lt (
\frac{\vp_\nu}{\ga_*}\rt )^2\rt\}\frac{d\vp_\nu}{\ga_*} . $$

Accomplishing the averaging (97), we will take into account the existence of 
the small parameters (61) and (62). The basic formulae that are met in the 
course of averaging the right--hand sides of (58) and (59) are assembled in 
the Appendix. Averaging the coupling functions in (78), we have
$$ \al \equiv \lgl\lgl\al_1\rgl\rgl = \al_0\lt (\frac{\ga_3}{\om_0}\rt )
\frac{\pi(\ga_2-\ga_3)^2}{(\ga_2-\ga_3)^2+\Dt^2} , $$
\be
\beta\equiv \lgl\lgl\al_2\rgl\rgl = \al_0\lt (\frac{\ga_3}{\om_0}\rt )
\frac{\pi(\ga_2-\ga_3)\Dt}{(\ga_2-\ga_3)^2+\Dt^2} .
\ee
The average effective frequency (82) and attenuation (83) are, respectively,
$$ \Om\equiv\lgl\lgl\Om_\vp\rgl\rgl =\om_0(1+\beta z) , $$
\be
\Ga\equiv \lgl\lgl\Ga_\vp\rgl\rgl =\ga_2 +\al\om_0z,
\ee
where an expansion in powers of the small parameters in (98) is used.

To write the evolution equations for the slow variables (96) in a compact 
form, we shall use some notation. Introduce the effective coupling parameter
\be
g \equiv\al\frac{\om_0}{\ga_2} =\al_0\lt (\frac{\ga_3}{\ga_2}\rt ) 
\frac{\pi(\ga_2-\ga_3)^2}{(\ga_2-\ga_3)^2+\Dt} .
\ee
Define the damping
\be
\ga_s \equiv\frac{f_0\ga_3^2}{8\om_0}\lt \{ x_0 +2\pi y_0 +
\frac{2\om_0z}{\Dt^2+\ga_2^2}\lt [ x_0(\beta\Dt-\al\ga_2) + 
y_0(\al\Dt +\beta\ga_2)\rt ]\rt\}
\ee
appearing when calculating the correlator $\;\lgl\lgl u_sh_f\rgl\rgl\;$ for 
the fields from (87) and (95), and also the attenuation
$$ \ga_f=\frac{f_0^2\ga_3^4}{32\om_0^2(\Dt^2+\ga_2^2)}\lt \{ \lt ( 1 +
\frac{8\pi^2}{3}\rt )\ga_2 - 2\pi\Dt +\right. $$
\be
\left. +\frac{\om_0z}{\Dt^2+\ga_2^2}\lt [ (\al-2\pi\beta)(\Dt^2-\ga_2^2) +
2\ga_2\Dt(\beta+2\pi\al)\rt ]\rt\}
\ee
resulting from the calculation of the correlator $\;\lgl\lgl u_fh_f\rgl\rgl\;$
for the fields (93) and (95).

Thus, the averaging of the right--hand sides of Eqs.(58) and (59), in 
compliance with (97), leads to the equations
$$ \frac{dz}{dt} =g\ga_2w -\ga_s -\ga_1(z-\zeta)-\ga_fz , $$
\be
\frac{dw}{dt}=-2\ga_2w -2(g\ga_2w -\ga_s)z +2\ga_fz^2
\ee
for the slow variables, where
\be
w \equiv v^2-2\ep_*z; \qquad \ep_*\equiv\frac{\ga_*^2}{\om_0^2} .
\ee

The quantities (101) and (102) characterize the relaxation of the 
magnetization owing to the action of the resonator field (95) formed by 
driving force (89). Note that $\;\ga_s\equiv 0\;$ for the incoherent initial 
condition, when $\;u_0 \equiv x_0-iy_0=0\;$. The squared amplitude of the 
driving force (89), remembering (44), can be written as
\be
f_0^2=\frac{8\al_0E_0^2}{\hbar\ga_3^2RN} .
\ee
This shows that $\;f_0\sim 1/\sqrt{N}\;$. Consequently, for the attenuations 
(101) and (102) we have $\;\ga_s\sim 1/\sqrt{N}\;$ and $\;\ga_f\sim 1/N\;$. 
These values for a macroscopic sample with $\;N\sim 10^{23}\;$ should be 
negligibly small.

In particular, if the electromotive force (88) corresponds to a resonance 
mode of the thermal Nyquist noise of the resonator, then [3] for its amplitude
we have
\be
E_0^2 =\frac{\hbar\om}{2\pi}\ga_3R{\rm coth}\frac{\hbar\om}{2k_BT} ,
\ee
where $\;k_B\;$ is the Boltzmann constant and $\;T\;$, temperature. For 
$\;\om\;$ in the radiofrequency region, typical of spin systems, (106) 
simplifies to
\be
E_0^2\simeq \frac{\ga_3}{\pi}Rk_BT \qquad \lt ( \frac{\hbar\om}{k_BT}
\ll 1\rt ).
\ee
Whence, for the amplitude in (105) we get
\be
f_0^2 =\frac{8\al_0k_BT}{\pi\hbar\ga_3N} \qquad (Nyquist \; noise) .
\ee
Substituting (108) into (101) and (102), we again come to the conclusion that 
these attenuations for a macroscopic sample are negligible. We shall exemplify
this by numerical estimates in Sec.X.

The conclusion that the radiation field of the coil does not provide a 
microscopic relaxation mechanism, so that $\;\ga_s\;$ and $\;\ga_f\;$ can be 
neglected in the equations for slow variables, is in complete agreement with 
the statement of Bloembergen and Pound [19] that a homogeneous magnetic field,
such as exists in the coil, will never produce the initial thermal relaxation 
in a macroscopic sample.

Let us acknowledge that $\;\ga_s\;$ and $\;\ga_f\;$ are negligibly small as 
compared to $\;\ga_2\;$. In addition, at low temperatures, characteristic of 
experiments [10-15], the spin--lattice damping is also much smaller than the 
spin--spin dephasing parameter. Thus, we have
\be
\frac{\ga_s}{\ga_2} \ll 1, \qquad \frac{\ga_f}{\ga_2} \ll 1, \qquad
\frac{\ga_1}{\ga_2} \ll 1 .
\ee
Taking into consideration (109), the slow--variable equations in (103) can be 
contracted to 
$$ \frac{dz}{dt}=g\ga_2w , $$
\be
\frac{dw}{dt}=-2\ga_2w(1 +gz) .
\ee

The  equations in (110) can be solved exactly in the following way. Notice, 
that the effective attenuation (99), with notation (100), acquires the form
\be
\Ga =\ga_2(1+gz) .
\ee
Using (111) in (110), we obtain
\be
\frac{d\Ga}{dt} =(g\ga_2)^2w, \qquad \frac{dw}{dt} = -2\Ga w .
\ee
Differentiating the first equation in (112), we come to
$$ \frac{d^2\Ga}{dt^2} +2\Ga\frac{d\Ga}{dt} = 0 , $$
which yields
\be
\frac{d\Ga}{dt} +\Ga^2 =\ga_0^2 ,
\ee
where $\;\ga_0\;$ is an integration constant. Eq.(113) is the Riccati equation 
whose solution is
\be
\Ga=\ga_0{\rm tanh}\lt (\frac{t-t_0}{\tau_0}\rt ) \qquad \lt ( \tau_0 \equiv
\frac{1}{\ga_0}\rt ) ,
\ee
where $\;t_0\;$, having the meaning of a delay time, is another integration 
constant. From (111) and (114) we have
\be
z =\frac{\ga_0}{g\ga_2}{\rm tanh}\lt (\frac{t-t_0}{\tau_0}\rt ) -
\frac{1}{g} ,
\ee
and from the first equation in (110) we find
\be
w =\lt (\frac{\ga_0}{g\ga_2}\rt )^2{\rm sech}^2\lt ( \frac{t-t_0}{\tau_0}
\rt ) .
\ee
The functions (115) and (116) are the exact solutions of (110). For the slow 
variable $\;v\;$, the relation (104) gives
\be
v^2 =\lt (\frac{\ga_0}{g\ga_2}\rt )^2{\rm sech}^2\lt (\frac{t-t_0}{\tau_0}
\rt )
+2\ep_*z .
\ee
As is seen, $\;\tau_0\;$ is an effective relaxation time.

The integration constants $\;\ga_0\;$ and $\;t_0\;$ are to be found from the 
initial conditions
\be
z(0)=z_0,\qquad v(0)=v_0 .
\ee
From (115), (117) and (118) we obtain
$$ \ga_0^2 =\Ga_0^2 +(g\ga_2)^2(v_0^2-2\ep_*z_0) , $$
\be
\Ga_0 \equiv \ga_2(1+gz_0); \qquad \ga_0\tau_0 = 1 ,
\ee
and the delay time
\be
t_0 =\frac{\tau_0}{2}\ln \lt |\frac{\ga_0 -\Ga_0}{\ga_0+\Ga_0}\rt | .
\ee
So, all constants in the solutions (115) and (117) for the slow variables are 
defined. The corresponding solutions for the fast variables are obtained by 
substituting (115) and (117) into the sums
$$ u=u_s + u_\vp + u_f , \qquad h = h_s + h_f , $$
whose terms  are given by (87),(84),(93),(94), and (95).

\vspace{3mm}

{\Large{\bf IX. Relaxation Regimes}}

\vspace{1mm}

Depending on the initial conditions and system parameters, one can distinguish
several qualitatively different relaxation regimes. The advantage of dealing 
with analytical solutions, as compared to numerical solutions, is that there 
are explicit formulas allowing direct investigation. When the problem contains
many parameters, as in the considered case, the detailed analysis of the 
solutions by varying the numerous parameters becomes excessively laborous 
if not impossible. At the same time it may happen, that not all parameters 
are equally  important, but only some of them or some their combinations. A 
striking example of this kind is presented by the problem considered here. 
Really, despite of great number of various parameters, characterizing the 
spin system coupled with a resonator, the solutions of evolution equations 
contain only several constants, the main of which is the effective coupling 
parameter (100). The general qualitative classification of different 
relaxation regimes can be done by varying only three quantities: the coupling 
parameter $\;g\;$, the initial polarization $\;z_0\;$, and the initial 
transverse magnetization $\;v_0\;$. The latter defines the level of initial 
coherence imposed on the system.

First of all, one can easily observe that if there is neither initial 
polarization, nor initial coherence, than (110) has only the trivial 
solution
\be
z=v=0 \qquad (z_0=v_0=0) .
\ee
Therefore, the necessary and sufficient condition for the existence of 
nontrivial solutions is a nonzero initial magnetization,
\be
m_0^2\equiv z_0^2 +v_0^2 > 0 .
\ee

The relation between the effective relaxation time $\;\tau_0\;$ and the 
spin--spin dephasing time $\;T_2\;$ depends on the value of $\;gm_0\;$. 
Namely,
$$ \tau_0 \approx T_2 \qquad (gm_0 \leq 1) , $$
\be
\tau_0 < T_2 \qquad (gm_0 > 1) ,
\ee
which follows from (119) under the assumption that $\;g\ep_*\ll 1\;$. The 
latter inequality is justified owing to the definition of $\;\ep_*\;$ in 
(104) as of a small parameter of second order with respect to (61).

The delay time (120) can have either negative or positive sign depending on 
the value of $\;gz_0\;$:
$$ t_0 \leq 0 \qquad (gz_0 \geq -1) , $$
\be
t_0 > 0 \qquad (gz_0 < -1 ) .
\ee
If $\;t_0 \leq 0\;$, then the maximum of the transverse magnetization (117) 
occurs at $\;t=0\;$. In this case, since $\;gz_0\geq -1\;$, then 
$\;\Ga_0>0\;$, which means that the amplitude of the fast variable $\;u\;$ 
decreases with time. When $\;t_0>0\;$, then the maximum of (117), i.e. the 
maximum of coherence, occurs at $\;t = t_0\;$. In this situation, as far as 
$\;gz_0<-1\;$, we have $\;\Ga_0<0\;$, which leads, according to (112), to the 
increase of the amplitude of $\;v\;$. The negative sign of the attenuation 
$\;\Ga_0\;$ means that the system acts as a generator.

Varying the quantities $\;gz_0\;$ and $\;gv_0\;$, we may distinguish seven 
qualitatively different relaxation regimes.

{\bf 1.}{\it Free induction}:
$$ g|z_0| < 1, \qquad 0<gv_0<1; $$
\be t_0 < 0 , \qquad \tau_0\approx T_2 .
\ee
This is the standard case of free nuclear induction, with the maximal 
coherence imposed at $\;t=0\;$ and relaxation time $\;T_2\;$. The coupling 
with a resonator plays no principal role. Note that the conditions of the 
upper line and lower line in (125) are not independent, but one line follows 
from another, in compliance with (123) and (124). However, we write down the 
relations between effective parameters, as well as those between 
characteristic times, to make the classification more physically 
transparent.

{\bf 2.}{\it Collective induction}:
$$ gz_0 > -1, \qquad gv_0 > 1; $$
\be
t_0 <0, \qquad \tau_0<T_2 .
\ee 
This case differs from the free induction by an essential role of the 
coupling with the resonator, which is sufficiently strong to develop 
collective effects leading to the shortening of the relaxation time 
$\;\tau_0\;$. When $\;gv_0\gg1\;$,then $\;\tau_0\ll T_2\;$. But, as in the 
previous case, the maximal coherence is that which is imposed at 
$\;t=0\;$.

{\bf 3.}{\it Free relaxation}:
$$ g|z_0|<1, \qquad v_0 =0; $$
\be
t_0<0, \qquad \tau_0 \approx T_2 .
\ee
The initial polarization $\;z_0\;$ and the coupling parameter $\;g\;$ are not 
sufficiently high for the appearance of self--organized coherence. At the 
same time, there is no imposed coherence. The relaxation process is mainly 
incoherent being due to the local random fields.

{\bf 4.}{\it Collective relaxation}:
$$ gz_0>1, \qquad v_0 =0 ; $$
\be
t_0 < 0, \qquad \tau_0 < T_2.
\ee
The difference with the previous case is that the positive initial 
polarization and the coupling parameter now are high, so that collective 
effects shorten the relaxation time. However, the initial state is close 
to a stationary one, and the change of $\;v\;$, being again due to the 
local fields, is too small to yield a noticeable coherence.

{\bf 5.}{\it Weak superradiance}:
$$ -2 < gz_0 < -1 , \qquad v_0 =0; $$
\be
t_0 > 0 , \qquad \tau_0 \approx T_2 .
\ee
The negative initial polarization corresponds to an inverted system. The 
value of this polarization and that of the coupling parameter $\;g\;$ are 
sufficient to make the delay time positive and to develop a weak coherence, 
as a result of incipient self--organization. But the latter is not yet enough 
strong to shorten the relaxation time.

{\bf 6.}{\it Pure superradiance}:
$$ gz_0 < -2 , \qquad v_0 =0 ; $$
\be
t_0 > 0 , \qquad \tau_0 < T_2 .
\ee
The system is prepared in a strongly nonequilibrium state with a high 
negative polarization. The coupling with a resonator is also strong. No 
initial coherence is imposed on the system. The coherence arises as a 
purely self--organized process started by local stochastic fields and 
developed owing to the resonator feedback field.

{\bf 7.}{\it Triggered superradiance}:
$$ gz_0 < -1, \qquad gv_0 > 1 ; $$ 
\be
t_0 > 0 , \qquad \tau_0 < T_2 .
\ee
The initial polarization is negative and the coupling with a resonator is 
strong enough, so that the collective behavior of spins, tight with each 
other through the feedback field, is important. But the relaxation is 
triggered by an imposed initial coherence. Therefore, this is a collective 
but not purely self--organized process.

In this classification, three regimes, free induction, collective induction, 
and triggered superradiance are triggered by an initial coherence thrust upon 
spins, that is by setting $\;v_0\neq 0\;$. Local random fields do not play an 
important role. Such kind of regimes can be described by the Bloch equations. 
Other four relaxation regimes, free relaxation, collective relaxation, weak 
superradiance, and pure superradiance, are initiated solely by local fields. 
No initial coherence is involved, i.e. $\;v_0=0\;$. The Bloch equations cannot
treat these four regimes.

Organizing the above classification, we separated qualitatively different 
relaxation types. As is clear, there can be intermediate kinds of relaxation 
in between these regimes. For example, the case when
$$ gz_0< -1, \qquad 0 < gv_0 < 1 $$
is between weak superradiance and triggered superradiance. In principle, 
everywhere in this classification the condition $\;v_0 =0\;$ can be replaced 
by $\;gv_0 < 1\;$, to include the intermediate regimes. However, it seems 
reasonable to distinguish, first, different physical reasons causing 
different relaxation mechanisms.

In the process of relaxation, the polarization (115), starting at 
$\;z=z_0\;$, tends to
\be
z \simeq \frac{\ga_0}{g}\lt ( T_2 -\tau_0\rt ) \qquad (t \gg t_0) .
\ee
If the initial polarization $\;z_0\;$ is negative, then (132) shows that a 
noticeable polarization reversal to a positive value occurs for the case 
when $\;\tau_0 < T_2\;$, that is for pure and triggered superradiance; also, 
it may happen at collective induction, though then the initial polarization 
is not high. The highest initial polarization is needed for pure 
superradiance. The corresponding polarization threshold is twice as large 
as that for weak superradiance or triggered superradiance. Eq.(132) shows 
as well that there can be no essential reversal of polarization from positive 
to negative values.  

It is illustrative to consider more in detail two limiting situations, when 
the coupling of the spin system with a resonator is either weak or strong. 
Start with the weak coupling limit, $\;g\ll 1\;$. Then for the relaxation 
width and relaxation time, from (119), we get
$$ \ga_0 \simeq \ga_2\lt [ 1 + gz_0 +
\frac{g^2}{2}\lt ( v_0^2 -2\ep_*z_0\rt )\rt ] , $$
\be
\tau_0 \simeq T_2\lt [ 1 -gz_0 -
\frac{g^2}{2}\lt ( v_0^2 -2z_0^2 -2\ep_*z_0\rt )\rt ] .
\ee
For the delay time (120) we have
\be
t_0 \simeq 
\frac{\tau_0}{2}\ln \lt | \frac{g^2}{4}\lt ( v_0^2 -2\ep_*z_0\rt )\rt | .
\ee
The behavior of polarization is
\be
z\simeq z_0 +\frac{g}{2}\lt ( v_0^2 -2\ep_*z_0\rt )\lt ( 1 - 
e^{-2\ga_2t}\rt ) .
\ee
When $\;g|z_0|<1\;$ and $\;gv_0<1\;$, we have the case of free induction 
(125), if $\;v_0\neq 0\;$. And if $\;v_0 =0\;$, then we have free relaxation 
(127) with 
$$ t_0 \simeq \frac{\tau_0}{2}\ln\lt |\frac{g^2}{2}\ep_*z_0\rt | , $$
\be
z\simeq z_0 -g\ep_*z_0\lt ( 1 - e^{-2\ga_2t}\rt ) .
\ee
The latter regime is entirely due to local fields, since if $\;\ep_*\;$ would 
be zero, then $\;z\simeq z_0\;$ and there would be no relaxation.

In the strong coupling limit, $\;g\gg 1\;$, from (119) we find
$$ \ga_0 \simeq \ga_2 \lt ( g\sqrt{m_0^2-2\ep_*z_0} +
\frac{z_0}{\sqrt{m_0^2 -2\ep_*z_0}}\rt ) . $$
This, using the inequality $\;\ep_* \ll 1\;$, can be reduced to
$$ \ga_0 \simeq gm_0\ga_2\lt [ 1 +\frac{z_0}{gm_0^2} -
\lt ( 1 -\frac{z_0}{gm_0^2}\rt ) \frac{\ep_*z_0}{m_0^2}\rt ] , $$
\be
\tau_0 \simeq \frac{T_2}{gm_0}\lt [ 1 - \frac{z_0}{gm_0^2} +
\lt ( 1 -\frac{3z_0}{gm_0^2}\rt )\frac{\ep_*z_0}{m_0^2}\rt ] .
\ee
The delay time (120) takes the form
\be
t_0 \simeq \frac{\tau_0}{2}\ln\lt |
\frac{m_0^2(m_0-z_0)(gm_0-1)-(gm_0^2-z_0)\ep_*z_0}{m_0^2(m_0+z_0)(gm_0+1) -
(gm_0^2-z_0)\ep_*z_0}\rt | .
\ee
For the final polarization (132) at $\;t \gg t_0\;$ we obtain
\be
z\simeq m_0 -\frac{1}{g}\lt [ 1 -\frac{z_0}{m_0} + 
\lt ( 1 -\frac{z_0}{gm_0^2}\rt ) g\ep_*z_0\rt ] .
\ee
These formulas for $\;gv_0 > 1\;$, depending on the value of $\;gz_0\;$, 
correspond either to collective induction (126) or to triggered superradiance 
(131). When $\;v_0=0\;$, we come, again depending on the value of $\;gz_0\;$, 
to collective relaxation,

(128), weak superradiance (129) or pure superradiance (130).

Note that if $\;gz_0 <-1\;$, then for any $\;v_0\;$ the maximal coherence is 
reached at $\;t=t_0 > 0\;$, when
\be
z(t_0) \approx -\frac{1}{g} , \qquad v(t_0)\simeq m_0 .
\ee

To better emphasize the role of local fields, let us analyse the case when 
there is no initial coherence, that is
\be
m_0=|z_0|, \qquad v_0 =0 ,
\ee
and $\;g|z_0|>1\;$. Then
$$ \ga_0 \simeq g|z_0|\ga_2\lt [ 1 +\frac{1}{gz_0} -\lt ( 1 -\frac{1}{gz_0}
\rt )\frac{\ep_*}{z_0}\rt ] , $$
\be
\tau_0 \simeq \frac{T_2}{g|z_0|}\lt [ 1 -\frac{1}{gz_0} +
\lt ( 1 -\frac{3}{gz_0}\rt )\frac{\ep_*}{z_0}\rt ] .
\ee
The delay time (120) becomes
\be
t_0 \simeq \frac{\tau_0}{2}\ln\lt |
\frac{(|z_0|-z_0)(g|z_0|-1) - (gz_0 -1)\ep_*}{(|z_0|+z_0)(g|z_0|+1) -
(gz_0-1)\ep_*}\rt | .
\ee
The final polarization (132) at $\;t \gg t_0\;$ is
\be
z\simeq |z_0| -\frac{1}{g|z_0|}\lt [ |z_0| -z_0 +(gz_0 -1)\ep_*\rt ] .
\ee

Consider separately the cases of positive and negative initial polarizations. 
When the latter is positive, i.e.
\be
z_0 =|z_0| ,
\ee
then the delay time (143) and final polarization (144) are
$$ t_0 \simeq \frac{\tau_0}{2}\ln\lt | 
\frac{(gz_0-1)\ep_*}{2z_0(gz_0+1)-(gz_0-1)\ep_*}\rt | , $$
\be
z\simeq z_0 -\lt ( 1 -\frac{1}{gz_0}\rt )\ep_* \qquad (t\gg t_0) .
\ee
Simplifying this for asymptotically large $\;gz_0 \gg 1\;$, and keeping in 
mind that $\;\ep_*\ll 1\;$, we have
$$ t_0 \simeq \frac{\tau_0}{2}\ln \lt |\frac{\ep_*}{2z_0}\rt | , \qquad 
\tau_0 \simeq \frac{T_2}{gz_0} , $$
\be
z \simeq z_0 -\ep_* \qquad (t \gg t_0) .
\ee
Formulas (146) and (147) correspond to collective relaxation (128) due to 
local fields.

Pass to the case of the negative initial polarization
\be
z_0 =-|z_0| .
\ee
Then, for the delay time (143) and final polarization (144) we find
$$ t_0\simeq\frac{\tau_0}{2}
\ln\lt |\frac{2|z_0|(g|z_0|-1)+(g|z_0|+1)\ep_*}{(g|z_0|+1)\ep_*}\rt| , $$
\be
z\simeq |z_0| -\frac{2}{g} +\lt ( 1 + \frac{1}{g|z_0|}\rt )\ep_* \qquad
(t \gg t_0) .
\ee
This can describe weak superradiance (129) or pure superradiance (130). Under 
the inequalities $\;g|z_0|\gg 1\;$ and $\;\ep_*\ll 1\;$ the latter expressions
change to
$$ t_0\simeq \frac{\tau_0}{2}\ln \lt |\frac{2z_0}{\ep_*}\rt | , \qquad 
\tau_0\simeq \frac{T_2}{g|z_0|} , $$
\be
z\simeq |z_0| -\frac{2}{g} +\ep_* \qquad (t\ll t_0) ,
\ee
which corresponds to pure superradiance (130). The origin of this phenomenon 
is completely due to local fluctuating fields.

An interesting question is: which part of dipole interaction is mainly 
responsible for starting the relaxation process in the regime of pure 
superradiance? Looking at Eqs.(57) and (58), we see that it is the random 
field $\;\vp\;$ which initiates the process, while $\;\vp_0\;$ only shifts 
the oscillation frequency. The stochastic field $\;\vp\;$ represents the 
local fields (54), which are related to the terms $\;b_{ij}\;$ and 
$\;c_{ij}\;$ of the dipole interactions (30). These terms are called 
nonsecular dipole interactions contrary to $\;a_{ij}\;$ that is called the 
secular dipole interaction [2]. In this way, it is the nonsecular dipole 
interactions that originate an initial relaxation and, consequently, the 
pure spin superradiance.

The obtained results make it possible to give one more justification for the 
term spin superradiance. For a system of $\;N\;$ nuclei an effective number 
of radiators may be defined as
$$ N_{eff} \equiv \frac{m_0N}{S} , $$
where $\;m_0\;$ is the initial magnetization introduced in (122) and $\;S\;$ 
is nuclear spin. Averaging the power of current
$$ P_\vp(t) \equiv Rj^2 =N\frac{\hbar\ga_3^2}{4\al_0}h^2 , $$
according to (97), we have
$$ P(t) \equiv\lgl\lgl P_\vp (t)\rgl\rgl = 
N(\al^2+\beta^2)\frac{\hbar\ga_2^2}{\al_0}v^2 . $$
The average current power for a superradiant regime has a maximum at 
$\;t =t_0 > 0\;$, where $\;v(t_0)=m_0\;$, is compliance with (140). 
Therefore,
$$ P(t_0) \sim m_0^2 \sim N_{eff}^2 . $$
Also, as is seen from (137), the radiation time
$$ \tau_0\sim m_0^{-1} \sim N^{-1}_{eff} . $$
The situation when the radiation pulse is proportional to the number of 
radiators squared, and the radiation time is inversely proportional to this 
number, is characteristic of superradiance.

Note that the intensity of magnetodipole radiation $\;I(t)\;$, as a function 
of time, behaves similarly to the current power $\;P(t)\;$ but contains a 
small factor making $\;I(t) \ll P(t) \;$, so that $\;P(t)\;$ is much easier 
to measure [17,19].

\vspace{3mm}

{\Large{\bf X. Numerical Estimates}}

\vspace{1mm}

The aim of the present paper is not to discuss some particular experiments 
but rather to give the general picture of possible relaxation processes. 
Nevertheless, the general qualitative picture can be better understood if 
illustrated by quantitative estimates. For this purpose, let us accept the 
values of parameters typical of experiments [11-15] with proton--rich 
materials, such as propanadiol $\;C_3H_8O_2\;$, butanol $\;C_4H_9OH\;$, and 
ammonia $\;NH_3\;$. Imploying the method of dynamic nuclear polarization, it 
is possible to polarize spins to a level of polarization reaching almost 
$\;100\%\;$. The samples polarized in this way are good examples of 
metastable nuclear magnets. The lifetime of such metastable materials at low 
temperature is very long. This time, $\;T_1\;$, is related to spin--lattice 
relaxation time. The order of its magnitude is given by the relation 
$\;T_1\sim(a/\Dt l)^2T_2\;$, in which $\;a\sim 10^{-8}cm\;$ is mean distance 
between spins, $\;\Dt l\sim 10^{-5}a\sim10^{-13}cm\;$ is the coefficient of 
linear magnetostriction, and $\;T_2\;$ is the spin--spin relaxation time. 
Whence, $\;T_1/T_2\sim 10^{10}\;$.

The spin--spin relaxation time is characterized by dipole interactions 
yielding $\;T_2\sim\hbar a^3/\mu^2\sim 10^{-5}s\;$. Consequently, 
$\;T_1\sim 10^5s\;$. The relaxation time $\;T_2^*\;$, related to local spin 
fluctuations, is also due to dipole interactions because of which 
$\;T_2^*\sim 10^{-5}s\;$.

In principle, there exists another longitudinal relaxation time due to the 
interaction of spins through the common electromagnetic field formed under 
the magnetodipole spin radiation. This time, which will be denoted by 
$\;T_1'\;$, to distinguish it from the spin--lattice relaxation time 
$\;T_1\;$, can be estimated as  $\;T_1'\sim (\lambda/a)^2T_2\;$, where 
$\;\lambda\;$ is the radiation wavelength. For the external magnetic field 
$\;H_0\sim 10^4G\;$, spins radiate in the radiofrequency region with 
$\;\om_0\sim 10^8s^{-1}\;$, thus with the wavelength $\;\lambda\sim 10^2cm\;$.
This gives $\;T_1'/T_2\sim 10^{20}\;$ or $\;T_1'\sim 10^{15}s\;$. As far as 
$\;T_1'/T_1\sim 10^{10}\;$, the longitudinal relaxation is practically due to 
the spin--lattice interactions only. The interaction through the radiation 
electromagnetic field is so weak, as compared to dipole interactions, that 
it does not play any role. This drastically distinguishes spin systems from 
atomic and molecular ones exhibiting superradiance. In the latter systems, 
the effective interaction through the common radiation field is not only 
important but serves as the basic mechanism for the appearance of strong 
collective correlations and coherence.

The resonator ringing time $\;T_3\;$ in the case of quasiresonance, when 
$\;\om\sim\om_0\sim10^8s^{-1}\;$, and for the quality factor $\;Q\sim 10^2\;$ 
is $\;T_3\sim 10^{-6}s\;$. The time of fast oscillations, defined in (64), is 
$\;T_0\sim 10^{-8}s\;$; so it is really the shortest among other 
characteristic times.

The damping parameters corresponding to the characteristic times in (63) are 
$\;\ga_1\sim 10^{-5}s^{-1}, \; \ga_2\sim 10^5s^{-1},\; \ga_2^*\sim 10^5s^{-1},
\;\ga_3\sim 10^6s^{-1}\;$. In this way, for the small parameters in (61) we 
have
$$ \frac{\ga_1}{\om_0}\sim 10^{-13}, \qquad \frac{\ga_2}{\om_0}\sim 10^{-3} ,
\qquad\frac{\ga_2^*}{\om_0}\sim 10^{-3}, \qquad
\frac{\ga_3}{\om}\sim 10^{-2}. $$

The coupling constant (44), owing to the relations 
$\;\hbar\ga_2\sim \mu^2/a^3\;$ and $\;\rho a^3 =1\;$, where $\;\rho\;$ is 
the particle density, is $\;\al_0\sim\pi\eta\ga_2/\ga_3\sim 10^{-1}\;$. 
The  average coupling functions in (98) are $\;\al\sim\ga_2/\om_0\sim 
10^{-3}\;$ and $\;\beta\leq \ga_2/\om_0\sim 10^{-3}\;$. In the case of 
exact resonance, when $\;\Dt=0\;$, the latter is identically zero, 
$\;\beta\equiv 0\;$. Thus, $\;\al\;$ and $\;\beta\;$ are also small 
parameters.

The maximal value of the effective coupling parameter (100) is of the order 
of $\;\pi^2\;$. Therefore it varies in the interval $\;0\leq g\propto 10\;$.

Consider the dampings (101) and (102) caused by the action of the 
electromotive force corresponding to a resonance mode of the thermal Nyquist 
noise with the amplitude (106). The typical temperature in experiments [11-15]
is $\;T\sim 0.1K\;$. As far as $\;k_BT\sim 10^{-5}eV\;$ and 
$\;\hbar\om\sim 10^{-7}eV\;$, we have $\;\hbar\om/k_BT\sim 10^{-2}\;$, 
hence the approximation (107) is justified. Using $\;\hbar\ga_3\sim
10^{-9}eV\;$, for the forcing--field amplitude (108) we find 
$\;f_0\sim10^2/\sqrt{N}\;$. Then, for the damping (101) we get 
$\;\ga_s\propto (10^5/\sqrt{N})s^{-1}\;$. In the case of passive initial 
conditions, when $\;x_0=y_0=0\;$, the value of (101) is exactly zero, 
$\;\ga_s\equiv 0\;$. Expression (102) yields $\;\ga_f\sim(10^7/N)s^{-1}\;$. 
For a sample of about $\;1cm^3\;$ the number of protons is 
$\;N\sim 10^{23}\;$. Thence, the thermal--noise forcing field has the 
amplitude $\;f_0\sim10^{-10}\;$, so for the damping (101) and (102) we get 
$\;\ga_s\leq 10^{-7}s^{-1}\;$ and $\;\ga_f\sim 10^{-16}s^{-1}\;$. These 
quantities are so much less than $\;\ga_2\;$ that there is no any reason to 
keep them in the equations. This also concerns $\;\ga_1\;$. Really, the 
relations in (109) are
$$ \frac{\ga_s}{\ga_2}\sim 10^{-12}, \qquad \frac{\ga_f}{\ga_2}\sim 
10^{-21}, \qquad \frac{\ga_1}{\ga_2}\sim 10^{-10} . $$
Therefore, the thermal Nyquist noise of a resonator has no influence on the 
spin dynamics in a microscopic sample. 

One might ask a question: What should be the size of a sample on which the 
resonator thermal noise could produce a noticeable effect? This would happen 
if $\;\ga_s\sim\ga_2\;$, which gives $\;N\sim 1\;$, or when 
$\;\ga_f\sim \ga_2\;$, from where $\;N\sim 100\;$. For $\;N > 100\;$ the 
Nyquist noise is practically of no importance.

The method of solving the equations, used in the present paper, makes it 
possible to take into account the retardation effects, related to the 
appearance of factors like $\;(1-e^{-\ga_3t})\;$. These effects are important 
for the correct description of relaxation processes. For example, the 
threshold of initial polarization for superradiance, weak or triggered, as 
follows from (129) and (131), is $\;z_0\sim-1/g\;$. In percentage, for spin 
$\;1/2\;$ and $\;g\sim 20\;$, this means that the superradiance threshold is 
$\;-10\%\;$. Respectively, the threshold of pure superradiance, given in 
(130), is $\;-20\%\;$. These values are in agreement with experiments 
[11-15]. While, if we would neglect the retardation replacing the factor 
$\;(1-e^{-\ga_3t})\;$ by $\;1\;$, then for the superradiance threshold we 
would get $\;-\ga_2/\al_0\om_0 = -\pi\ga_3/g\om_0\sim 10^{-3}\;$. In 
percentage, this makes $\;-0.1\%\;$, which is unrealistically small.

In the regime of pure spin superradiance, the characteristic times 
$\;\tau_0\;$ and $\;t_0\;$ can be estimated from (150). Since 
$\;\tau_0\sim T_2/g|z_0|\;$, taking $\;g|z_0|\sim 10\;$, we find the 
radiation time $\;\tau_0\sim 10^{-6}s\;$. The local--field parameter, 
defined in (104), is $\;\ep_*\sim 10^{-6}\;$. Whence, for the delay time 
we obtain $\;t_0\sim(3\div 5)\tau_0\;$, that is 
$\;t_0\sim 10^{-6}-10^{-5}s\;$. The reversed final polarization, 
according to (150), can reach $\;90\%\;$. Note that the problem of the 
fast polarization reversal of proton solid--state targets is of great 
practical importance for the study of scattering in high and intermediate 
energy physics [15]. The phenomenon of spin superradiance can be used to 
achieve the desired fast repolarization.

\vspace{3mm}

{\bf Acknowledgements}

\vspace{3mm}

It is a pleasure for me to express my deep gratitude to R.Friedberg, 
S.R.Hartmann, and J.T.Manassah for fruitful discussions and useful comments. 
I am also grateful to J.L.Birman, S.R.Hartmann and J.T.Manassah for their 
advises, support and kind hospitality during my visits to the Columbia 
University and City University of New York, where these results have been 
reported. 

This work was partly supported by the Natural Sciences and Engineering 
Research Council of Canada.

\newpage

{\Large{\bf Appendix}}

\vspace{1mm}

Here we present the basic formulas for the averages defined in (97) and used 
in Sec.VIII when deriving the equations for slow variables.

For the stochastic fields, with the Gaussian distribution in mind, we have
$$ \lgl\lgl \vp_0\rgl\rgl = \lgl\lgl\vp_1\rgl\rgl =\lgl\lgl\vp_2\rgl
\rgl=0 , $$
$$ \lgl\lgl\vp_0^2\rgl\rgl =\lgl\lgl\vp_1^2\rgl\rgl =\lgl\lgl\vp_2^2\rgl
\rgl =\ga_*^2 . $$
Note that, instead of defining a particular distribution, we could postulate 
the above properties of random fields.

In the following expressions the averaging (97) is accompanied by expansions 
in powers of small parameters (61):
$$ \lgl\lgl e^{-\Ga_\vp t}\rgl\rgl \simeq 1 -\frac{\pi\Ga}{\om_0} , $$
$$ \lgl\lgl e^{i(\Om_\vp +i\Ga_\vp)t}\rgl\rgl \simeq 
\frac{\Om-\om_0 +i\Ga}{\om_0} , $$
$$ \lgl\lgl e^{(i\Dt -\ga_2)t}\rgl\rgl\simeq 
1 +\frac{\pi}{\om_0}(i\Dt -\ga_2),$$
$$ \lgl\lgl e^{i\om t}\lt (1 -e^{-\ga_3t}\rt)\rgl\rgl \simeq
-i\frac{\ga_3}{\om_0} , $$
$$ \lgl\lgl e^{i\Dt t}\lt (1 -e^{-\ga_3t}\rt )\rgl\rgl \simeq 
\pi\frac{\ga_3}{\om_0} , $$
$$ \lgl\lgl e^{i(\om+\Om_\vp+i\Ga_\vp)t}\lt ( 1 -e^{-\ga_3t}\rt )\rgl\rgl
\simeq -i\frac{\ga_3}{2\om_0} , $$
$$ \lgl\lgl \lt ( 1 -e^{-\ga_3t}\rt )^2\rgl\rgl \simeq 
\frac{4\pi^2\ga_3^2}{3\om_0^2} , $$
$$ \lgl\lgl e^{2i\om t}\lt ( 1 -e^{-\ga_3t}\rt )^2\rgl\rgl \simeq
\lt ( 1 -2\pi i\rt ) \frac{\ga_3^2}{2\om_0^2} , $$
where $\;\Om_\vp\;$ and $\;\Ga_\vp\;$ are given by (82) and (83), 
respectively, and $\;\Om\;$ with $\;\Ga\;$ are defined in (99). 

Emphasize the importance of the factor $\;(1-e^{-\ga_3t})\;$ responsible for 
the retardation effects.

\end{document}